\newcommand{\degree}{\ensuremath{^{\circ}}}

\documentclass[5p,twocolumn]{elsarticle} 

\usepackage{color}
\usepackage{lineno}
\usepackage{amssymb}
\usepackage{comment}
\usepackage{float}
\usepackage[normalem]{ulem}

\journal{Nuclear Instruments and Methods A}

\begin{document}
\begin{frontmatter}

\title{A gamma- and X-ray detector for cryogenic, high magnetic field applications}

\author[a]{R. L. Cooper}
    \address[a]{Indiana University, Bloomington, IN 47408 USA; roblcoop@indiana.edu}
\author[c]{R. Alarcon}
    \address[c]{Arizona State University, Tempe, AZ 85287 USA}
\author[d]{M. J. Bales}
    \address[d]{University of Michigan, Ann Arbor, MI 48109 USA}
\author[e]{C. D. Bass}
    \address[e]{National Institute of Standards and Technology, Gaithersburg, MD 20899 USA; thomas.gentile@nist.gov}
\author[b]{E. J. Beise}
\author[b]{H. Breuer}
    \address[b]{University of Maryland, College Park, MD 20742 USA; breuer@enp.umd.edu}
\author[g]{J. Byrne}
    \address[g]{University of Sussex, Brighton, BN1 9QH UK}
\author[d]{T. E. Chupp}
\author[f]{K. J. Coakley}
    \address[f]{National Institute of Standards and Technology, Boulder, CO 80305 USA}
\author[e]{M. S. Dewey}
\author[e]{C. Fu}
\author[e]{T. R. Gentile\corref{cor1}}
\author[e]{H. P. Mumm}
\author[e]{J. S. Nico}
\author[c]{B. O'Neill}
\author[i]{K. Pulliam}
    \address[i]{Tulane University, New Orleans, LA 70118 USA}
\author[e]{A. K. Thompson}
\author[i]{F. E. Wietfeldt}
  
\cortext[cor1]{Corresponding author:  Thomas R. Gentile, Stop 8461, NIST, Gaithersburg, MD 20899 USA; phone 301-975-5431; FAX 301-926-1604; email: thomas.gentile@nist.gov}
  
\begin{abstract}

As part of an experiment to measure the spectrum of photons emitted in 
beta-decay of the free neutron, we developed and operated a detector consisting
of 12 bismuth germanate (BGO) crystals coupled to avalanche photodiodes (APDs).
The detector was operated near liquid nitrogen temperature in the
bore of a superconducting magnet and registered photons with energies from 5 keV to 
1000 keV.  To enlarge the detection range,
we also directly detected soft X-rays with energies between 0.2 keV and 20 keV  
with three large area APDs.  The construction and operation of the detector
is presented, as well as information on operation of APDs at cryogenic temperatures.  

\end{abstract}

\date{today}


\end{frontmatter}

\section{Introduction}
\label{sec:intro}

A free neutron beta-decays to a proton, electron, and an antineutrino and is predicted to be accompanied by 
a spectrum of photons with energies up to 782 keV~\cite{Gaponov96a, Gaponov96b, 
Gaponov00, Bernard04a, Bernard04b}, (i.e.,~$n \rightarrow p + e^{-} +\overline{\nu} +\gamma$).
The photons arise primarily due to inner bremsstrahlung from the high energy electron.
Classically, we can think of this process as an electron at rest being rapidly accelerated
to its final velocity or imagine that charge is turned on in the same time interval~\cite{Jackson}.
The first observation of the photons accompanying neutron decay was reported (RDK I)~\cite{Nico06}
and a detailed analysis determined the branching ratio to be $(3.09 \pm 0.31) \times 10^{-3}$ 
for  photon energies between 15~keV and 340~keV~\cite{Cooper10}.  
This result is consistent with the theoretical value 
of $2.85 \times 10^{-3}$ for the same photon energy range~\cite{Gardner06}.  
The uncertainty in the branching ratio for RDK I was dominated by systematic uncertainties,
but these effects are believed to be controllable.  
Thus to perform a more accurate measurement of the spectrum,  higher counting rates are required. 
A second experiment with an improved detection system was recently performed (RDK II)
with the aim of measuring the radiative decay spectrum 
and branching ratio to $\approx$~1~\% relative uncertainty.  The broad improvements in the RDK II
experiment were outlined in a previous work~\cite{Cooper09}.  In this paper we describe the construction
and operation of the photon detection system in detail.

To provide the context for the description of the improved photon detector, we briefly describe 
the mode of operation of the experiment; further details can be found
in Refs.~\cite{Cooper10, Cooper09, Cooper,Gentile07}.  
The experiments were conducted on the NG-6 neutron beam line\cite{Dewey05} at the 
National Institute of Standards and Technology (NIST) Center for Neutron Research (NCNR) in Gaithersburg, Maryland.  
A collimated neutron beam (2.5 cm diameter for RDK II) passed along the axis of a 4.6~T superconducting solenoid magnet 
which is $\approx$~50~cm long.  
The high magnetic field ensures that when a neutron decays, the high-energy electron and recoil 
proton are confined to cyclotron orbits with radii less than 1~mm.  
At the upstream end of the apparatus, the magnetic field is bent by 9.5~$^{\circ}$ to guide the charged
particles out of the beam and towards a silicon surface barrier detector (SBD).  
(In this paper we use the terms  \textquotedblleft downstream\textquotedblright  and
\textquotedblleft upstream\textquotedblright\ 
to refer to the directions parallel and antiparallel to the neutron velocity vector.)

In the RDK I experiment, a single 12~mm by 12~mm by 200~mm bar of bismuth
 germanate (BGO) was coupled to a large-area avalanche photodiode (APD) and placed at the bottom 
of the magnet bore, along the length, and just outside of the neutron beam.  
The strong magnetic field inside the bore precluded the use of photomultiplier tubes (PMTs), whereas at the time we began this work APDs had been reported to be insensitive to magnetic fields~\cite{Boucher03}.
The detector operates near liquid nitrogen temperature, which substantially decreases the noise in the APD and increases
its gain, and increases the light output of the BGO crystals~\cite{Piltingsrud79, Gironnet08}.  
  
A candidate radiative decay event is identified in software by requiring the correlation of a prompt
electron signal in the SBD, a simultaneous photon signal in the photon detector,
and a delayed, low-energy proton pulse in the SBD.  Hence accurate determination of the  
timing of the photon signal with respect to the electron signal is important for rejection of  
background photons.  Digitizing the preamplifier pulses, rather than simply counting events  
using a spectroscopy amplifier with discrimination, allowed for optimized timing as well as off-line  
analysis of the photon signal waveforms after the experiment. 
The temporal resolution has been improved by about a factor of two compared to the results in Ref.~\cite{Gentile07}
by improved light collection, electronics, and signal-to-noise ratio.

To improve upon the RDK~I experiment, a twelve element detector was constructed
to increase the solid angle for photon detection.
In addition the range was extended  down to 0.2 keV using three larger APDs
for direct detection of soft X-rays.   In the course of developing this direct detector we found that the 
response of the APDs employed is strongly distorted when operated at cryogenic temperatures with
a strong magnetic field in the APD plane\cite{Gentile11}.  In Sec.~\ref{sec:photon} we describe the RDK II photon
detector and in Sec.~\ref{ssec:performance} we present its performance.  
Several features of operating 
the BGO crystals and APD detectors near liquid nitrogen temperature  are presented in Sec.~\ref{ssec:APD features}. 

\section{Photon Detection System}
\label{sec:photon}

In this section, we provide general motivations for the construction of the RDK II photon detector, 
followed by a detailed description of the construction.
  
\subsection{General considerations} \label{ssec:photon.general}

Due to the long neutron lifetime (15 min) and the relatively low intensity
of neutron beams, the detection rate of electron/delayed proton coincidences in the RDK II apparatus 
was only $\approx$~10~s$^{-1}$.
Due to the low branching ratio for neutron radiative decay,
a large solid angle for detection is desired.  The photon emission rate
increases rapidly with decreasing photon energy, so obtaining a low detection threshold is also desirable.
High resolution is not required because the neutron radiative decay produces a smooth, featureless
spectrum of photons similar to that of external bremsstrahlung.

\begin{figure} [p]
\begin{center}
\includegraphics[width=3.2in]{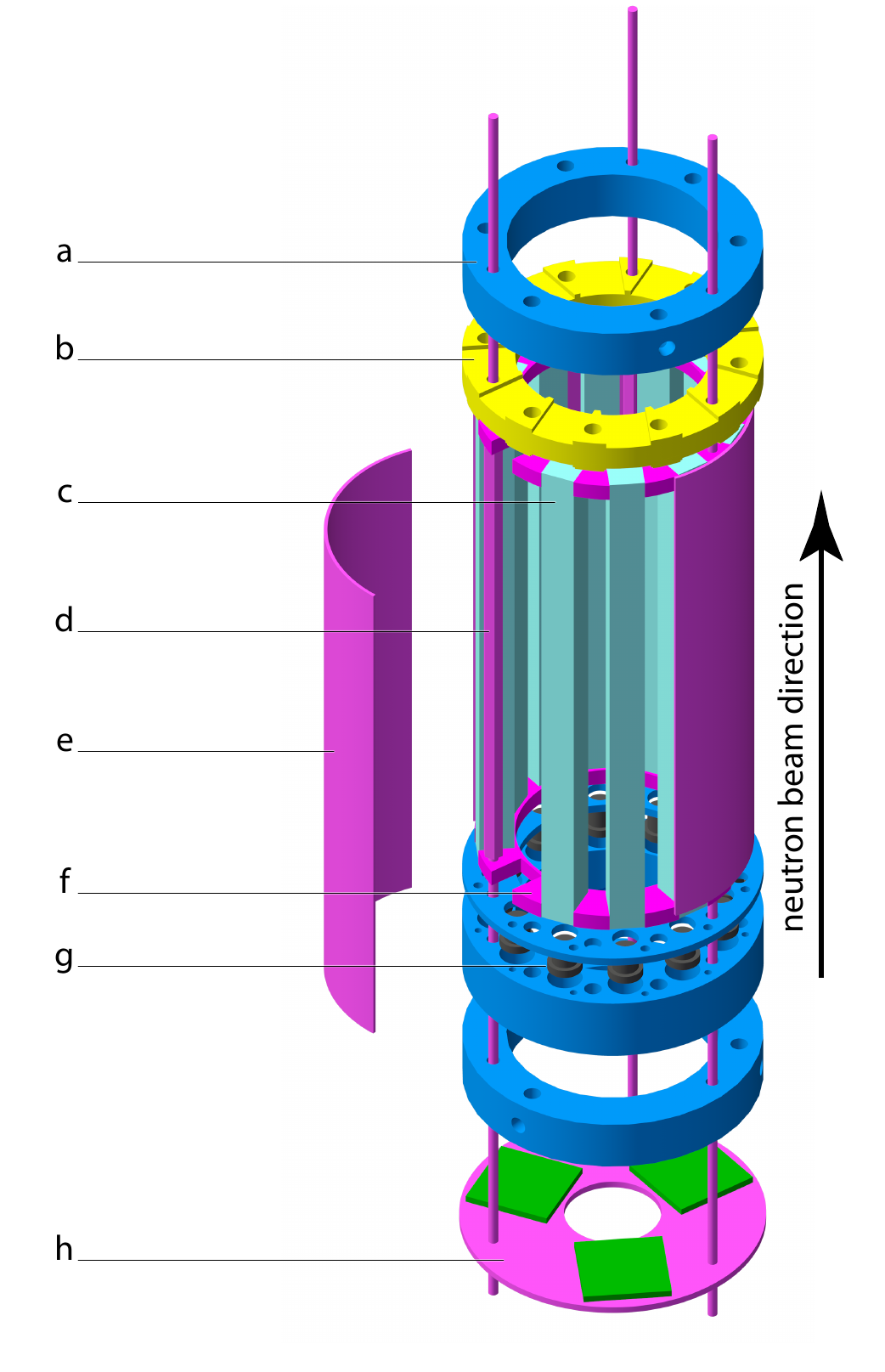}
	\caption{ A schematic view of the detector. 
(a) nylon rings (blue); (b) G10 APD holder (yellow); (c) BGO crystals (blue-green); (d) aluminum support rods (violet);
(e) aluminum heat shield (violet); (f) aluminum crystal holders (violet);  (g) nylon crystal pushers (black); 
(h) APD soft-X-ray detector (see Sec.~\ref{directAPD}) consisting of aluminum
support plate (violet) and large APDs (green).  The neutron beam (black arrow) passes through
the center entering from upstream (bottom) and exiting downstream (top).
The APD detectors (not shown) are held in bottom cutouts of the APD holder.
The  BGO crystals (one crystal is removed to show structure) are held
in cutouts in the crystal holder.  The spring loaded crystal pushers between the upstream nylon rings
push the BGO crystals against the APDs.  Support rods secure the structure along the beam axis.
The cylindrical heat shields secure the BGO crystals radially and provide heat shielding.\label{Alan_overview_new}}
\end{center}
\end{figure}

Whereas doped inorganic crystals generally exhibit lower light yield at 80~K relative to room temperature,
the yield from pure inorganic crystals increases at low temperature~\cite{Piltingsrud79, Gironnet08}.
As discussed elsewhere~\cite{Gentile07}, both pure CsI and BGO were investigated; BGO was chosen 
because pure CsI is hygroscopic, exhibited greater spatial dependence in the light yield, and
yields a higher probability of Compton scattering.  The RDK II detector consists of an annular array of twelve
12 mm by 12~mm by 200~mm~\cite{SaintGobain, Rexon}
BGO bars that surrounds the cold neutron beam.  To maximize the solid angle, we briefly considered a single annular BGO crystal coupled to APDs, but this results in substantially decreased light collection.  For the relatively low
gain of APDs compared with PMTs, such light loss is unacceptable.
For a 12 mm path length, BGO has nearly 100~\% absorption below 200 keV,
decreasing to 50~\% at 900 keV~\cite{SaintGobain}.  
The BGOs were placed in mechanical contact with large-area APDs with no additional optical coupling.  
The 13.5 mm $\times$ 13.5 mm active area of each APD~\cite{RMD} was slightly larger than
the end face of each BGO crystal to ensure full coverage of the crystal and
facilitate alignment between the APD and BGO crystal.  
To conserve physical space and avoid failure due to thermal stress upon cooling,
these APDs lacked the typical protective outer casing.
Nevertheless, the APDs proved to be rugged and did not suffer from direct contact with the hard 
surfaces of the BGO crystals.

\subsection{Mechanical Setup}
\label{ssec:mechanical}

\begin{figure} [p]
	\centering \includegraphics[width=3.0in]{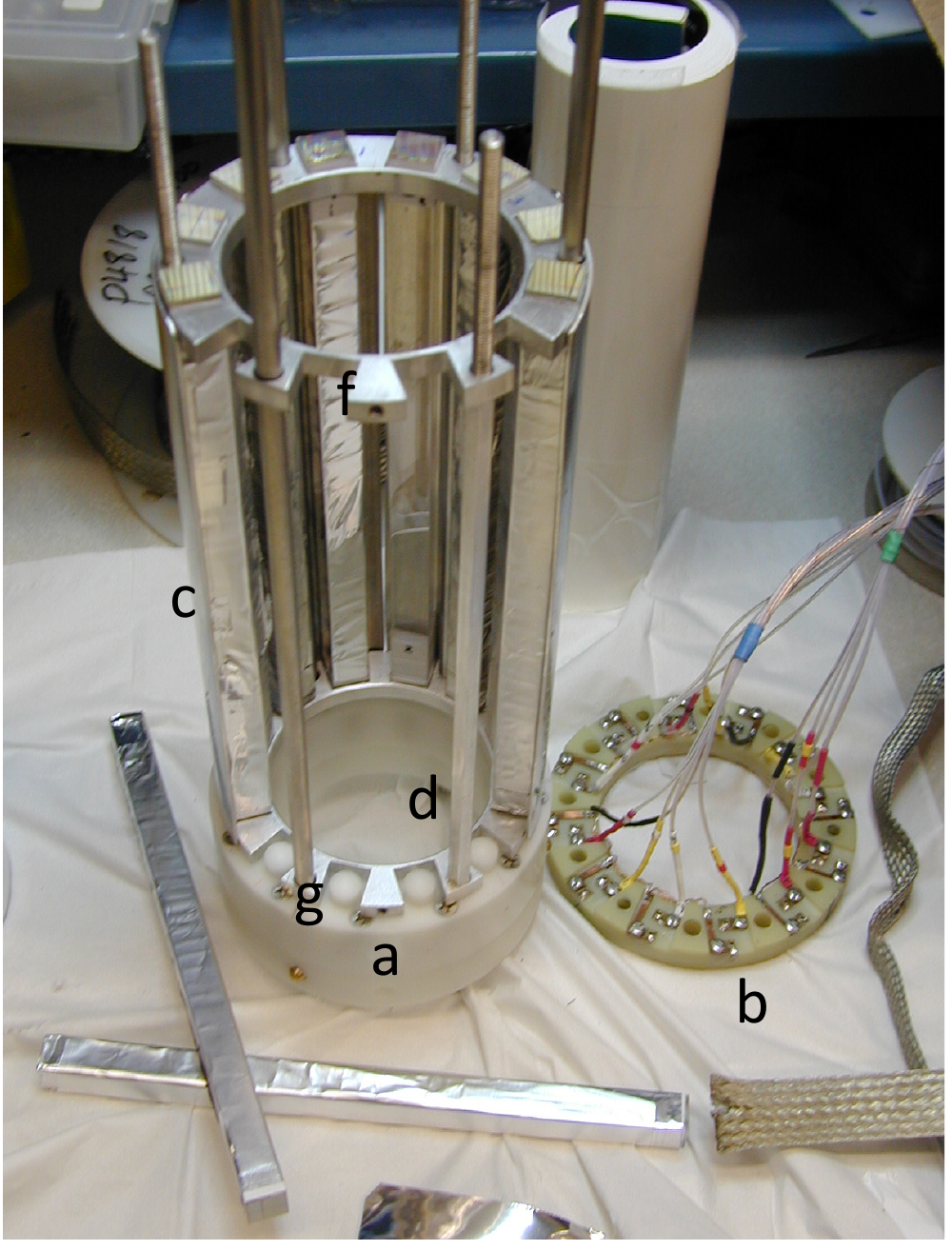}
\caption{Photograph of the partially assembled photon detector system.
See Fig.~\ref{Alan_overview_new} for labels provided.
Two  BGO crystals, partially wrapped in aluminized Mylar foil, are in the foreground.
At the bottom of the standing structure are the upstream nylon rings with one of the three brass
centering pins visible. Above the nylon ring are two crystal holders (f).
Three aluminum rods space the positioning rings and  three stainless steel rods move the brass centering pins.
The crystal pushers (g) are visible in the cutouts of the lower positioning ring. 
The G10 APD holder (b) with signal cables is visible on the right. 
\label{Photo_disassembled}}
\end{figure}

The BGO crystals and APD detectors are mounted in a low-mass, non-magnetic mechanical structure, which 
fits into the bore of the magnet.  A schematic 
view of the system is shown in Fig.~\ref{Alan_overview_new} and a photograph of the partially 
disassembled system is shown in Fig.~\ref{Photo_disassembled}.

The BGO crystals are positioned at 30~\degree\ intervals and at a radius of 36 mm (inside of crystals) by two aluminum rings
with cutouts to define their position.  The crystals are held in place lengthwise by non-metallic rings at both ends.
Spring-loaded pistons in the nylon ring (upstream) ensure that contact between the BGO crystals
and the APDs is preserved when the apparatus is cooled.  The G10 ring (downstream)
provides spaces for the APD detectors (see right side of Fig. \ref{Photo_disassembled}).
Three 120\degree\  sections of an aluminum cylinder (1.65 mm wall thickness)
constrain the crystals to their slots and are thermally connected to the magnet's liquid nitrogen reservoir.
Fig. \ref{Photo_disassembled} shows the APD holder ring on the right, with the APDs facing
away from the viewer and the electrical contacts with wiring visible.

The detector package is designed so that it can be radially adjusted
by up to 3 mm in order to center it with respect to the beam.  
Three brass adjusters with hemispherical heads (to reduce thermal
contact) were placed radially within each nylon ring, and each was
translated with a stainless steel rod.  Fig.~\ref{Photo_disassembled} shows a brass
foot extending from the bottom-left nylon ring with its corresponding adjustment rod. 

Four silicon diode temperature sensors~\cite{diodes} are used to monitor the temperatures at the upstream
and downstream ends of the detector package as well as on the liquid-nitrogen and
liquid-helium-cooled magnet surfaces.
To block background gamma rays from the reactor and possible
correlated bremsstrahlung radiation from the SBD,  a small Pb block was placed
just upstream of the single BGO bar used In the RDK I experiment.
In RDK II, a 2.5~cm thick Pb annulus was employed,
with 1.8~mm thick copper added to absorb secondary Pb X-rays (72 keV to 80 keV).

\begin{table}[p]
  \begin{tabular} {ccc}
    \hline
    far end & long & $F$ \\ 
    \hline
    -  & Al foil or Al-Mylar &  $<$1.05  \\
    Al foil  & - & 1.6 \\
    PTFE  & - & 1.8 \\
    paint & 10 mm painted & 2.0 \\ 
    paint & - & 2.2 \\ 
    roughened, paint & -  & 1.8  \\
    \hline
  \end{tabular}
  \caption {Summary of tests of BGO surface coverage to improve light collection efficiency.
   The \textquotedblleft far end\textquotedblright\ face is the 12 mm by 12 mm surface opposite from the PMT;
    the four  \textquotedblleft long\textquotedblright\ surfaces are perpendicular to the PMT and far end surfaces.
    The factor $F$ is the light yield relative to the case of no surface coverage;
    the reported values have a typical standard uncertainty of 0.05. \textquotedblleft Al-Mylar\textquotedblright\ - aluminized Mylar;  
    PTFE - polytetrafluoroethylene; \textquotedblleft paint\textquotedblright\  - diffuse reflective paint
    (Bicron BC-622~\cite{SaintGobain}); \textquotedblleft 
    roughened\textquotedblright\ - roughened with 600 grit sand paper
     (later \textquotedblleft re-polished\textquotedblright \ with very fine (12000 grit) paper).
    }
  \label{table:surfcov}
\end{table}

\subsection{Light Collection}
\label{ssec:signal}

\begin{figure} [p]
\centering
\includegraphics[scale=0.50]{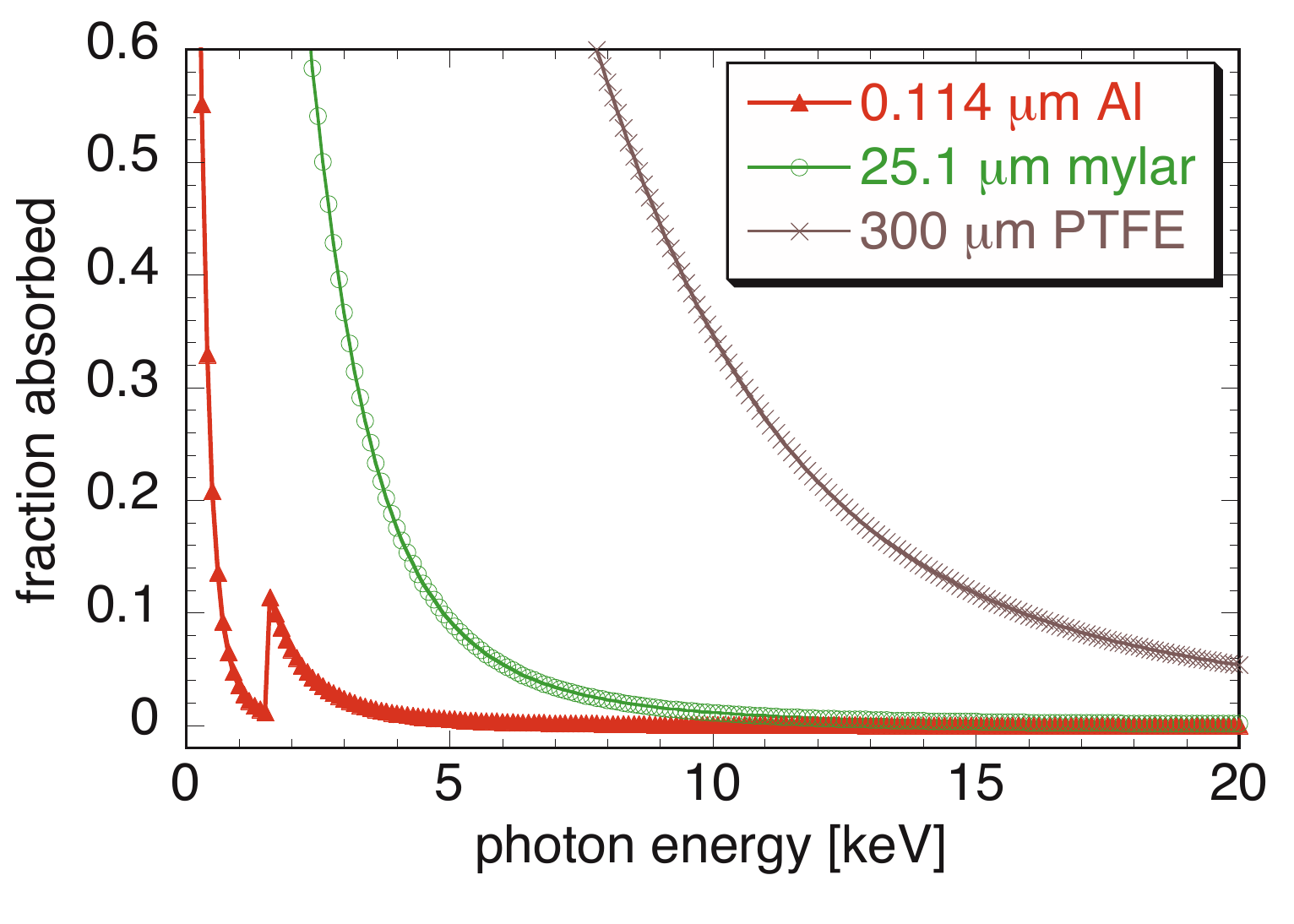}
\caption{Absorption curves for cover components of the BGO crystals as a function of
photon energy. Most of the crystal surfaces are covered only by aluminized Mylar 
(0.114~$\mu$m Al, red triangles;  25.1~$\mu$m Mylar, green circles). 
For protection, the end sections are also covered by
300~$\mu$m PTFE (brown crosses). The absorption data were adapted from~\cite{XrayAbs}.\label{BGO_COVER}}
\end{figure}

To maximize the BGO light output, we investigated
an optical coupler between the BGO crystal and the APD and
covering or treating the BGO surfaces to increase the amount of light that reaches the APD.
Light transmission between the BGO crystal and APD can be improved by providing a medium with an index
of refraction $n$ between $n$=1 and $n$=2.15 (BGO),
where we have assumed there is a microscopic gap between the BGO crystal and the APD.
Operation at cryogenic temperatures precluded the use of optical grease and other options,
but the use of silicone rubber sheet has been reported for temperatures as low as $-$20$~^\circ$C~\cite{Ikagawa05}.
We studied the coupling of BGO to APD with silicone rubber (GE RTV615; $n$=1.406~\cite{MGchemicals}).
According to Monte Carlo simulations, this optical coupler should increase the
amount of light extracted from the BGO crystal by a factor 2.3.  
However, we found that optical contact was lost for tests near liquid nitrogen temperature,
thus yielding no improvement in light collection. 
We expect that the difference in thermal contractions of the crystal, the hardened silicone rubber, and the APD is responsible for this loss of contact.  While it might be possible to find a suitable optical coupler,
 this method was not pursued any further. 
Thus the present detector has an n=1.0 gap between the BGO crystals and APDs.

\begin{figure} [p]
\centering
\includegraphics[scale=0.48]{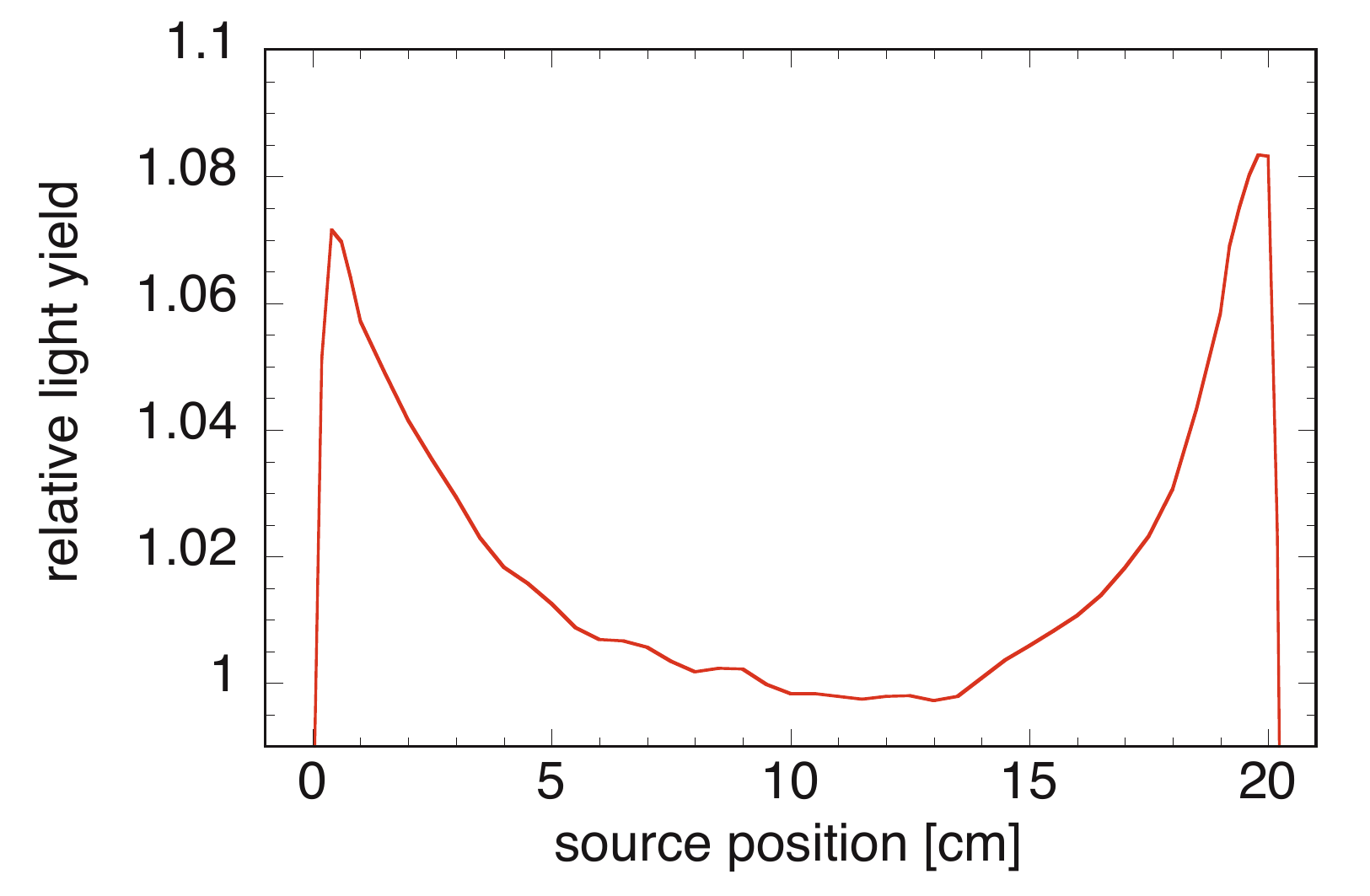}
\caption{Light yield detected in APDs as a function of source position along the BGOs,
relative to that obtained at the central source position of 10 cm.  The
124 keV line from a collimated $^{57}$Co source was moved from the APD (0 cm) to the 
painted end of the BGO crystals (20 cm). The curve represents the average normalized yield
of 11 detectors operated near liquid nitrogen temperature.\label{posdep}}
\end{figure}

Studies were made of different methods of covering or treating the BGO detector
surfaces to optimize light collection.  For convenience these tests were performed at room temperature
using the 662 keV gamma-ray from a $^{137}$Cs source to illuminate the
BGO crystal near its center. A PMT, separated from one of the
crystal surfaces by an air gap, was used to detect the scintillation signals.
Our results are summarized in Table~\ref{table:surfcov}.

The results are consistent with expectations from Ref.~\cite{Keil70}.
A diffuse reflector on the far end increases light collection by redirecting some otherwise
trapped light into the angular cone which allows light to escape the crystal towards the PMT.
Light which can escape towards the PMT internally reflects from the long surfaces, therefore applying
diffuse paint on those surfaces reduces light collection in the PMT. 
We decided to paint the far ends of each BGO crystal with reflective paint and to
loosely cover the remaining detector with reflective foil, mainly for
mechanical and external light-intrusion protection.
Specifically, the long sides of each BGO crystal were covered by two layers of aluminized 
Mylar (total of 12.4 $\mu$m Mylar and 0.065 $\mu$m aluminum). An
additional aluminized Mylar sheet (12.7 $\mu$m Mylar
and 0.049 $\mu$m aluminum) separated the beam from the BGO
crystals. The long-side regions near the end faces ($\approx$12 mm)
had additional layers of PTFE tape and
aluminized Mylar for mechanical protection from the metal positioning rings
($\approx$300 $\mu$m PTFE and 6 $\mu$m Mylar with 0.033 $\mu$m aluminum).

The covering on the BGOs was deliberately kept thin to reduce absorption of low 
energy photons and thus help minimize the energy detection threshold for the experiment.
Absorption curves for the cover materials used are shown in Fig. \ref{BGO_COVER}. As
shown, the absorption of aluminized Mylar and PTFE decrease to less than 5~\% above $\approx$~6.5 keV
and $\approx$~21~keV, respectively. 

Ref.~\cite{Keil70} also predicts for our configuration
(a photon detector coupled via an n=1 gap to the crystal, diffuse reflector on opposite end)
that the detected light yield for uniform light creation depends on the location along the length of
the crystal. Indeed we found that the detected light created with collimated sources 
increases at both ends, near the APD and near the far end (Fig. \ref{posdep}). 

During initial design we considered using APDs on both crystal ends, but rejected this approach due to increased cost and additional complexity in construction and data acquisition. We later found that diffusely reflecting paint on the far end yielded 
an improvement in light collection comparable to what would be expected for APDs at both ends.

\subsection{High voltage and signal processing}
\label{ssec:electrical}

The electrical ground and high voltage (HV) connections
(which also carry the APD signals) were provided by PTFE-tube
insulated 32 AWG (American Wire Gauge, 0.20 mm diameter) phosphor-bronze wires
within the magnet's vacuum chamber.  These thin wires were chosen over coaxial cables
to minimize heat load and outgassing. Although these signal/HV cables were not shielded and shared 
ground wires in groups of four, cross talk was found to be less than 0.5~\% of signal amplitude. 

Outside of the magnet, cabling continued via 30 cm long 
double-shielded RG-59 cables to Canberra 2006 preamplifiers~\cite{Canberra}. 
Double shielding the coaxial HV/signal cables between the vacuum
feedthrough and preamplifier reduced low-amplitude noise considerably.
The preamplifier output signals were directly fed into two 8-channel GaGe
cards~\cite{Gage}, which were set to digitize the signals with 40 ns/channel resolution
in 80 $\mu$s long traces.  The decay time constant of the preamplifiers is 50 $\mu$s.
Because the gain of the APDs is highly sensitive to the high voltage, 
we employed power supplies~\cite{NHQ} that could be set in 0.1 V steps ($\approx$5~\% gain change for 1 V change).
Details of gain dependencies are described in Sec.~\ref{subsec:gain}.
The APD gain was found to be stable to 1~\%, and gain drifts could be correlated
 with drift in the temperature as determined from the silicon diode temperature sensors.

\begin{figure}[p]
\begin{center}
\includegraphics[scale=0.18]{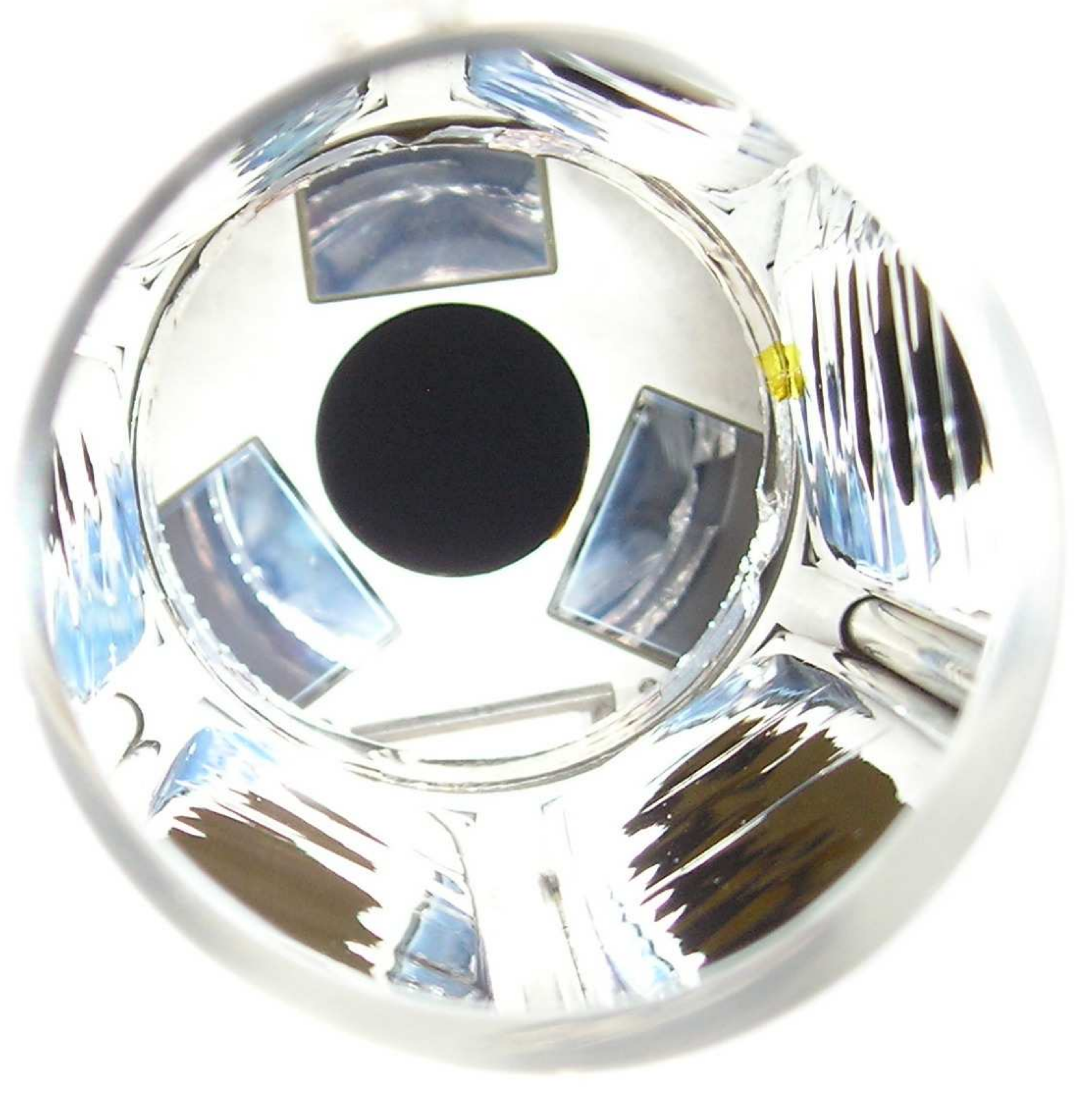}  
\caption{Photograph of the soft X-ray detector, looking upstream through
the scintillator detector towards the three 28 mm by 28 mm APDs.
The three APDs are partially occluded in this view and the
shiny cylindrical surface visible is the layer of aluminized
Mylar that separates neutron beam from BGO crystals.\label{fig:dirdetphoto} }
\end{center}
\end{figure}
\subsection{Direct soft X-ray detector}
\label{directAPD}

Soft X-rays from neutron radiative decay with energies between $\approx$0.2~keV
and 20 keV were directly detected with additional APDs~\cite{Cooper09}.  The detector, shown
in Fig.~\ref{fig:dirdetphoto}, consisted of three 28 mm by 28 mm (active area) APDs~\cite{RMD} mounted
around the neutron beam with the plane of the APD perpendicular to the
magnetic field.   This configuration was chosen because we found significant 
distortion of the APD response when the plane of the APD was parallel
to the magnetic field (and thus the APD electric field was perpendicular to the
magnetic field)~\cite{Gentile11}.   In the configuration shown in Fig.~\ref{fig:dirdetphoto}
the only effects of the magnetic field for tests with 5.9 keV X-rays provided by
an $^{55}$Fe source were a small (less than 10~\%) decrease in the gain and a decrease of
up to a factor of two in the width of the peak. A $^{55}$Fe source was mounted near the APDs to provide
a continuous energy calibration during the experiment.   The APDs were read out
with the same preamplifier and GaGe card configuration as for the
APDs in the scintillator detector.

\section{Performance}
\label{ssec:performance}

\subsection{BGO/APD scintillator detector}
\label{ssec:bgoapd}

\begin{figure}[p]
\includegraphics[scale=0.47]{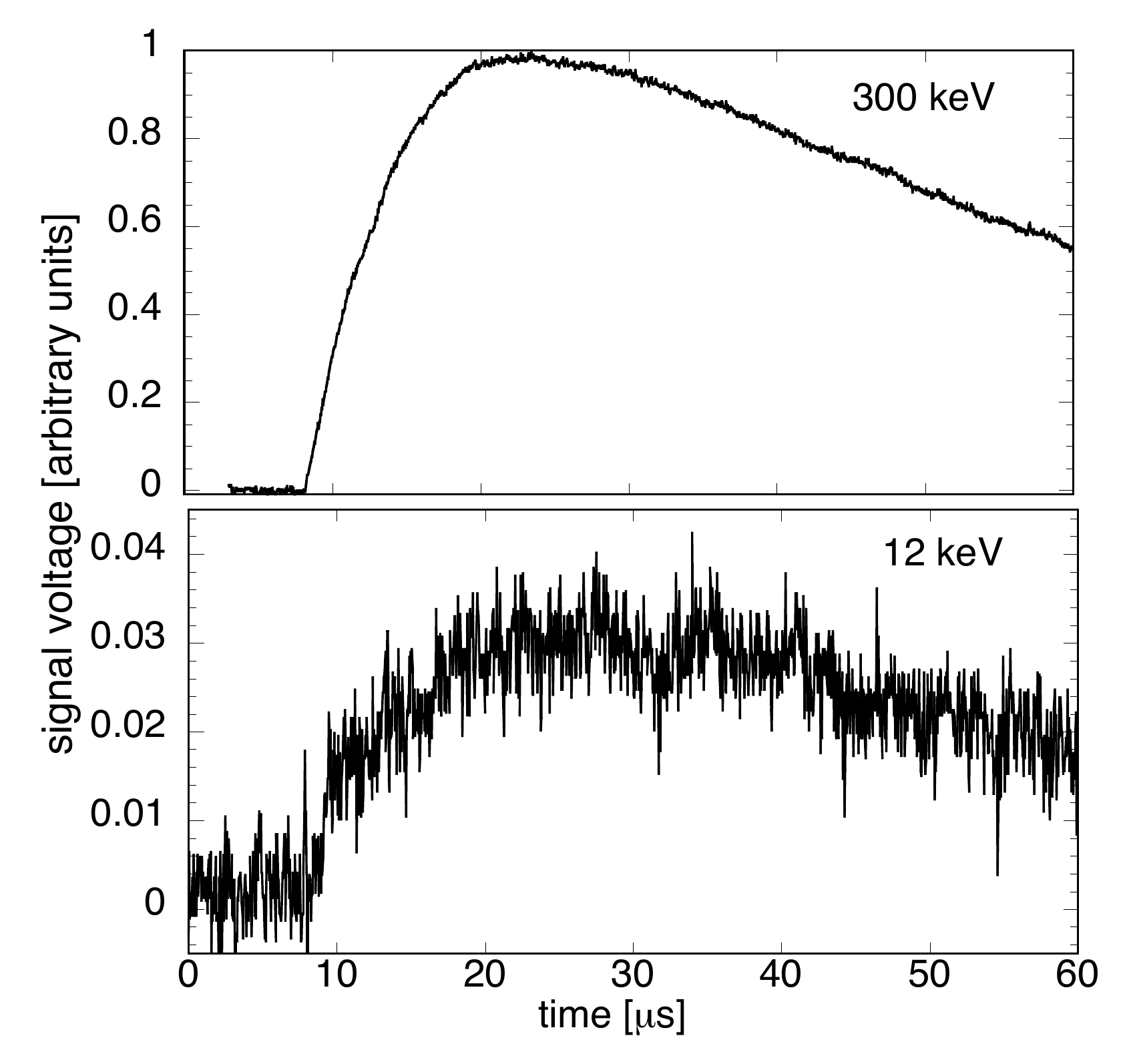} 
\caption{APD signals after the charge integrating preamplifier for individual 300 keV (top) and
12 keV (bottom) photons incident on the BGO crystal.  The 5~$\mu$s exponential rise time
determined from the 300 keV signal corresponds to the BGO scintillation decay time. \label{fig:signal_BGOevent} }
\end{figure}
 
Signals from the BGO/APD detector are shown in Fig.~\ref{fig:signal_BGOevent} for 300~keV and 12~keV photons
that were recorded during the radiative decay experiment; the energy
calibration was determined from 511 keV gamma-rays from electron-positron
annihilation that were present in the background spectrum.  The peak-to-peak noise  corresponds to 
a minimum detectable photon energy of $\approx$ 5~keV.
The 5~$\mu$s rise time is governed by the time scale for scintillation in the BGO crystal,
which increases substantially at low temperatures~\cite{Weber73}.

\begin{figure}[p]
\includegraphics[scale=0.48]{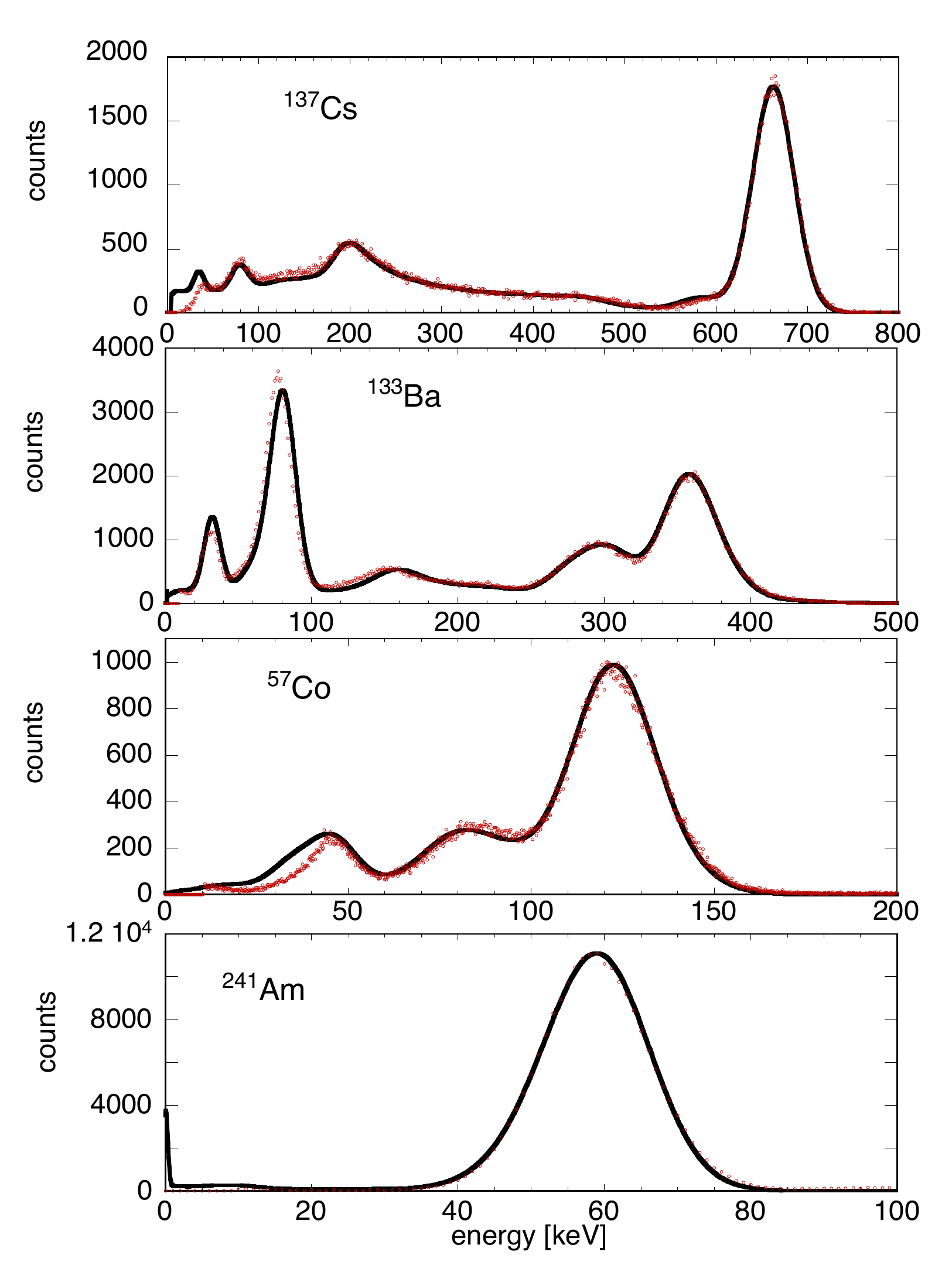}   
\caption{BGO/APD detector pulse height spectra for illumination by radioactive sources: 
$^{137}$Cs (662 keV and 32 keV to 37 keV X-rays), $^{133}$Ba (80 keV, 303 keV, 356 keV and 32 keV to 37 keV X-rays),
$^{57}$Co (primary line at 122 keV), and $^{241}$Am (59.5 keV). 
Data are shown by red markers and predicted spectra based on a model of the detector response are shown by solid black lines.
The calibration for the x-axis is the 662 keV peak from the $^{137}$Cs source.
For all sources except $^{241}$Am, bismuth X-ray fluorescence emitted by the
other 11 BGO crystals in the array (74 keV to 90 keV, primary line at 77 keV) is visible.
Besides the photopeak response, one also observes
a continuum of incomplete energy deposition due to Compton scattering.
For the $^{57}$Co source the peak at $\approx$45 keV is from incomplete energy deposition
due to escape of bismuth X-rays from the BGO crystal.\label{fig:AmCoBaCs_MC}}
\end{figure}

To test the detector response with radioactive sources, we employed a reentrant tube
that allowed the sources to be outside the magnet's vacuum chamber while illuminating
the detector in a similar way as the photons from neutron decay.
Pulse height spectra for tests of the BGO/APD detector with $^{137}$Cs, $^{133}$Ba, $^{57}$Co, and $^{241}$Am  sources
are shown in Fig.~\ref{fig:AmCoBaCs_MC}, along with the results of modelling
the detector response based, in part, on MCNP5~\cite{Brown03} predictions for the energy deposit spectrum
in each  BGO crystal.   The experimental results are being used to test this modelling,
which is required for analysis of the neutron radiative decay experiment.
A more detailed treatment will be described in a future publication.
Although nonlinearity in the response of BGO has been reported\cite{Dorenbos95,Mos04,Verdier11,Khodyuk} the modeling shown in Fig. 7 assumes a linear response.  We are currently investigating the magnitude of nonlinearity in the overall detection system for our experimental operating conditions.

The typical energy resolution is $\approx$~10~\% FWHM (full width at half maximum) at 662 keV and 
$\approx$~30~\% at 60 keV, dominated by the statistics of
electron-hole generation and multiplication in the APD~\cite{Mos03}.
The observed resolution is close to the value of 11~\% that we
estimate using a typical excess noise factor of 2.5~\cite{Moszynski02b,Ludhova05} and
$\approx$~1000 photoelectrons expected for a 662 keV gamma ray, where we have assumed
a photon yield for BGO of $\approx$~24 keV$^{-1}$~\cite{Mos04},  a photon collection efficiency of 10~\%  (based
on Ref.~\cite{Keil70} and Monte Carlo simulations), and a quantum efficiency
of 70~\%~\cite{McClish06} for the green BGO fluorescence (peak at $\approx$~530 nm at 90~K~\cite{Piltingsrud79}).

\subsection{Direct soft X-ray detection}
\label{ssec:dirdet}

\begin{figure}[p]
\includegraphics[scale=0.47]{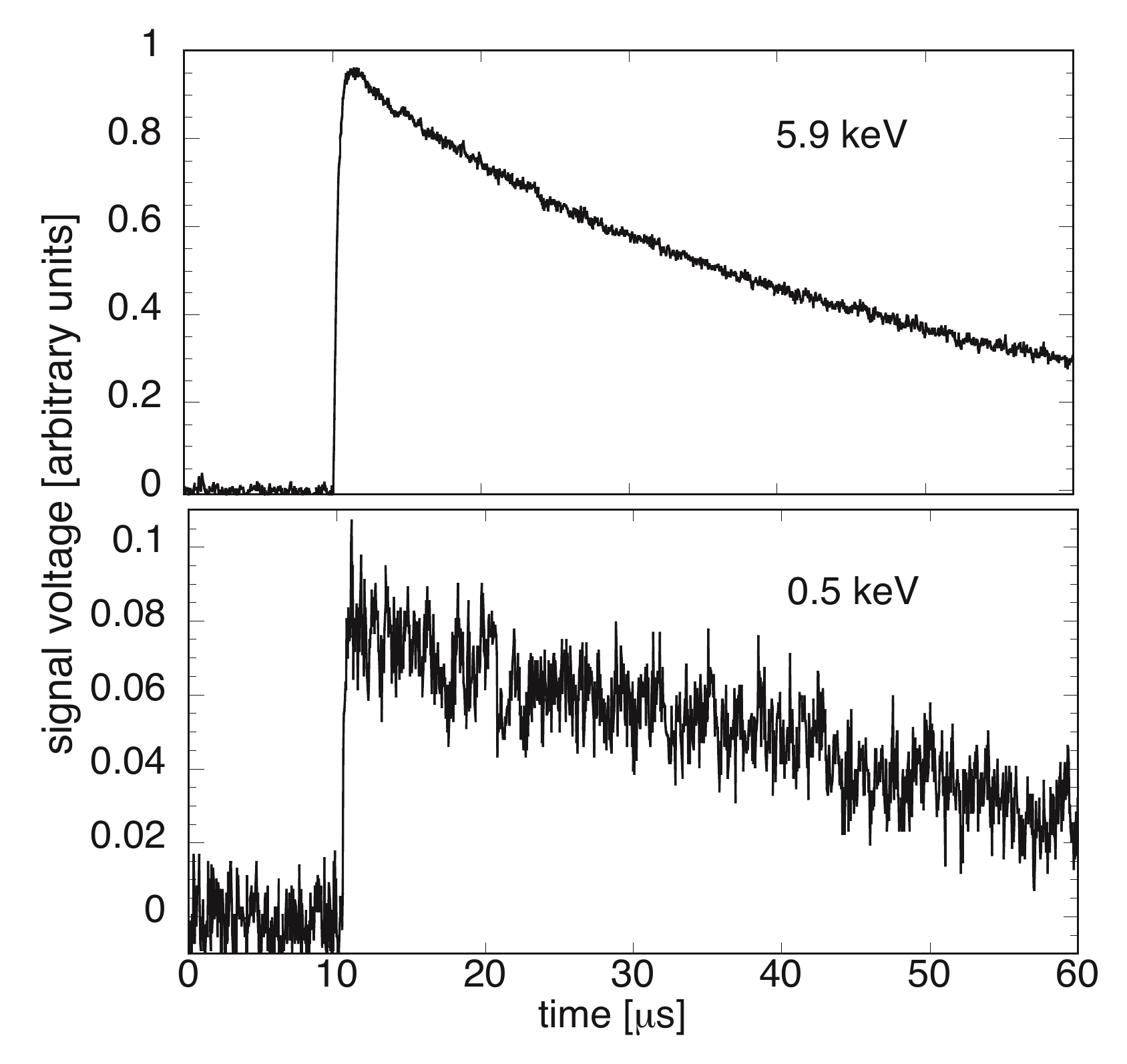}  
\caption{APD signals after the preamplifier for a 5.9 keV X-ray from the $^{55}$Fe source
and a signal with an amplitude corresponding to a 0.5 keV X-ray.  The exponential rise time 
determined from the 5.9 keV X-ray is 0.3~$\mu$s.\label{fig:APDevent} }
\end{figure}

\begin{figure}[p]
\includegraphics[scale=0.5]{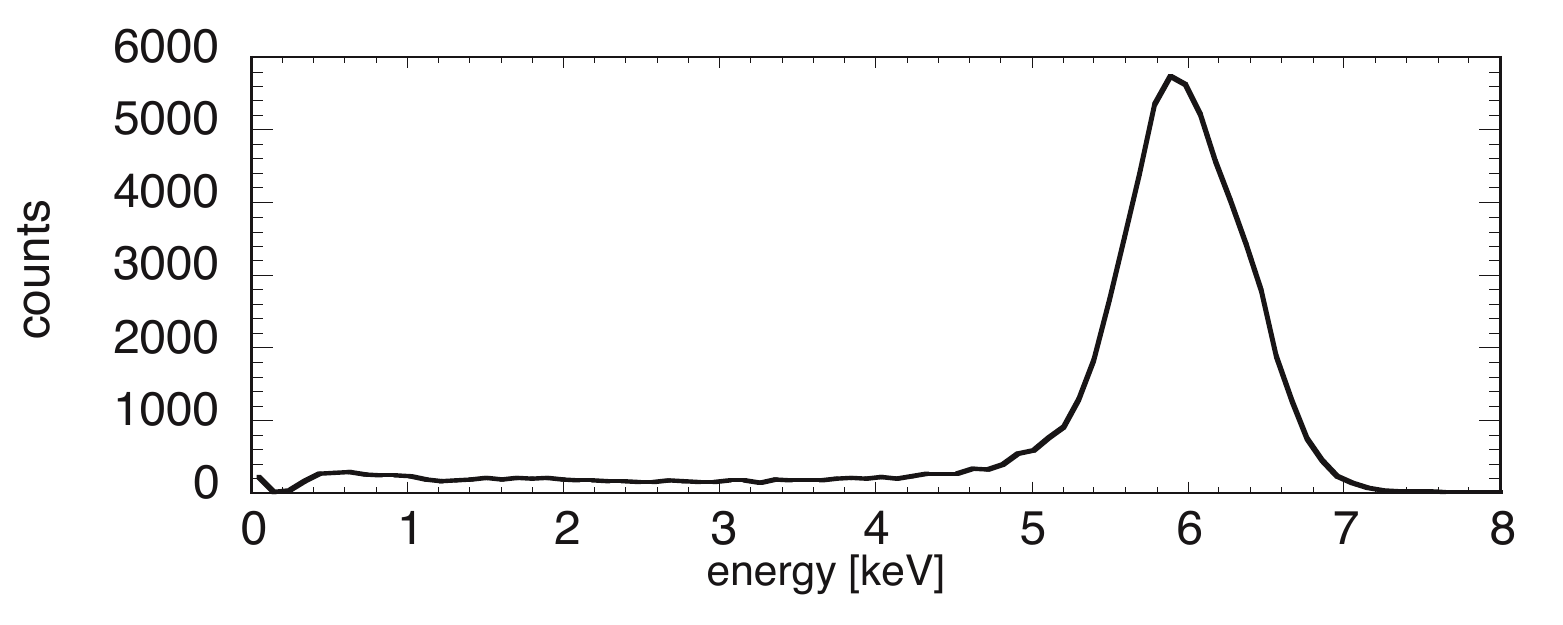}  
\caption{APD pulse height spectrum for  5.9~keV X-rays from an $^{55}$Fe source.\label{fig:Fe_new} }
\end{figure}

Fig.~\ref{fig:APDevent} shows APD signals after the preamplifier for a 5.9 keV X-ray from the $^{55}$Fe source
and a signal with a amplitude corresponding to a 0.5 keV X-ray; 
the latter was obtained during the neutron decay experiment.
The rise time of the signal from the soft X-ray detector is much faster than those
of the scintillator detector.  The peak-to-peak noise  corresponds to 
a minimum detectable photon energy of $\approx$ 0.2~keV.
Fig.~\ref{fig:Fe_new} shows the typical pulse height spectrum
from the APD for 5.9~keV X-rays from a  $^{55}$Fe source.
At high magnetic field the typical resolution for the response to 5.9 keV X-rays was 15~\% FWHM.

Between 1 keV and 5 keV, X-rays are detected by APDs with close to 100~\%
efficiency~\cite{Farrell91}. Above 5~keV, the detection efficiency decreases due to 
decreasing absorption by silicon.   For the 60~$\mu$m depletion layer for the APDs
employed, the detection efficiency decreases by an order of magnitude between 5~keV and 16~keV.
Primarily below 1~keV but also just above the Si K-edge at 1.8 keV, there are
changes in the pulse height spectrum arising from the short
penetration depth into the silicon coupled with strong doping at the front of the APD.
In addition, there is increased absorption from the protective glass layer on the APD.
The complex response in the 0.35~keV to 1.5~keV range has been studied by measurements
with monochromatic X-ray beams and synchrotron radiation~\cite{Gentile11a}.
As the X-ray energy is reduced below $\approx$1~keV, 
an increasing fraction of events that have incomplete collection
of photoelectrons was observed.  For two of the three APDs used in the neutron experiment,
the detection efficiency for our experimental threshold was found to decrease by an order of magnitude
between 1.0 keV and 0.4 keV.  For the third APD (from
a different wafer) the low energy response was improved, which we presume
is due primarily to a shallower depth of doping.

\section{Cryogenic Features of APDs}
\label{ssec:APD features}

Aspects of cryogenic operation of APDs have previously been reported~\cite{Moszynski02b,Yang03}.
During testing and operation of the detector at cryogenic temperatures,
we observed several features that were important for the performance of this experiment.
These features were studied in the BGO/APD detectors,
but should dominantly represent APD properties and not properties of the BGO crystals.
Although the features described here have only been explored for the APDs used~\cite{RMD},
they may serve as a guide of what to expect and what to look for
when operating APDs at cryogenic temperatures.

The gain of a given APD depends strongly on the magnitude of the applied high voltage ($V_{A}$) and temperature.  In addition we observed some dependence on magnetic field strength, as previously noted (see Sec.~\ref{directAPD}). Furthermore we observed a dependence on how much time has elapsed since the high voltage has been applied, and the size of this effect varied with temperature

\subsection {Operating high voltage }
\label{subsubsec:breakdown}

At low temperatures, e.g., near liquid nitrogen temperatures, one can
define a limiting \textquotedblleft breakdown high voltage\textquotedblright, $V_{bd}$,
within 1 V. At this voltage, noise signals (best observed without a radioactive source) increase in amplitude
so that they saturate a preamplifier. Within 3 V 
above $V_{bd}$ the bias current, normally below
1~nA, increases to more than 10 nA.  No damage to the APDs was observed even for $V_{A}$
up to 100 V above $V_{bd}$ for several seconds.
In this experiment $V_{bd}$ ranged from 1387~V to 1479~V for the twelve APDs operating at 86 K.

The gain-to-noise ratio has a broad maximum in the range
of about 20 V to 50 V below $V_{bd}$.  The absolute gain of
an APD at a given voltage below breakdown in this range depends on the individual APD;
for the 12 detectors of this experiment at 90~K the observed signal amplitudes for a given
photon energy absorbed in the BGOs varied by up to a factor of 2.1 at $\Delta V_{A} = -30$ V,
where $\Delta V_{A} = V_{A} - V_{bd}$ is the voltage below breakdown.

\subsection{Gain dependence}
\label{subsec:gain}

\begin{figure} [p]
\centering
\includegraphics[scale=.50]{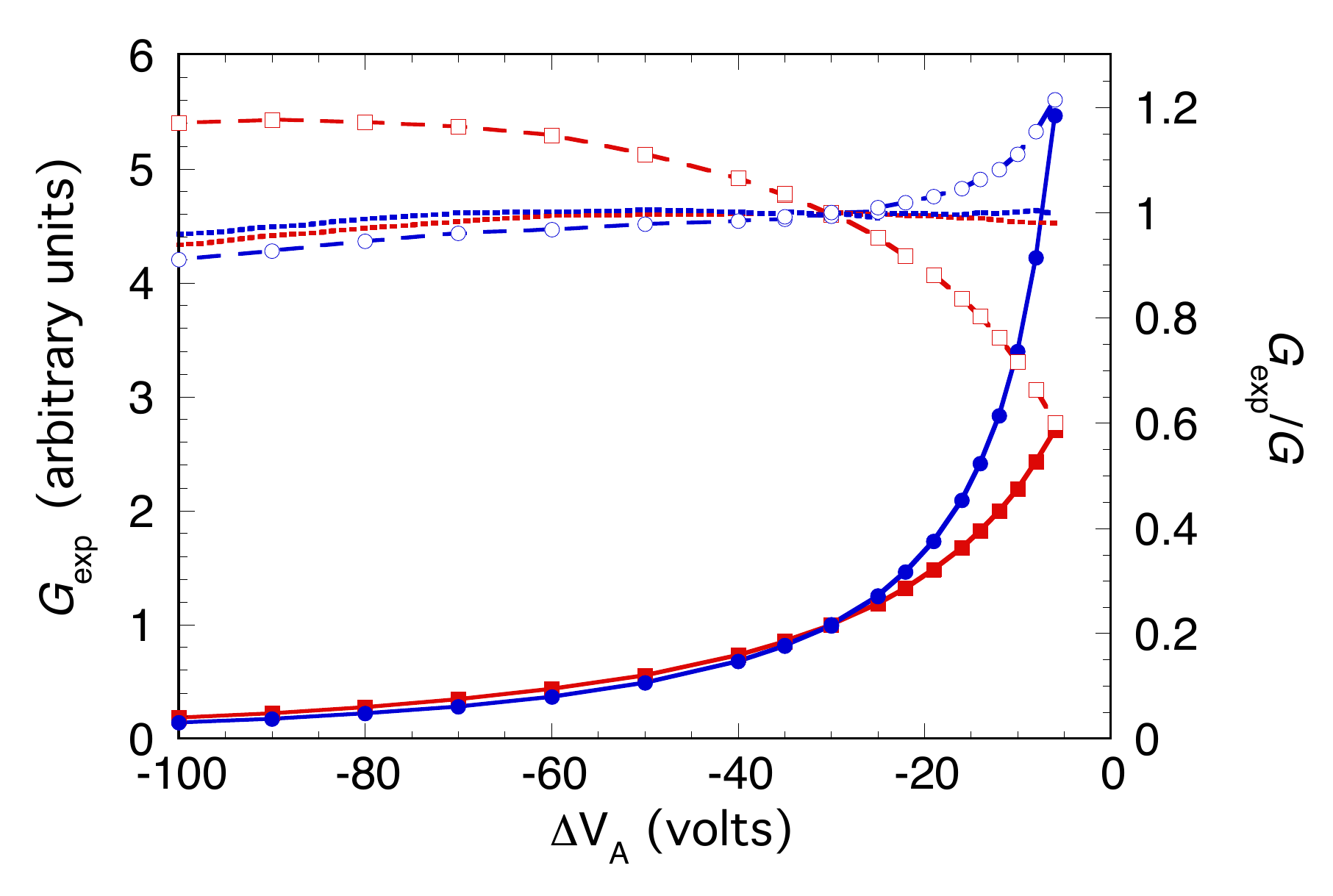}
\caption{Fig.~\ref{gain_hv}.   The gain $G_\mathrm{exp}$ of two APDs (filled symbols, left axis) as a function of $\Delta V_{A}$, normalized to 1.0 at $\Delta V_{A}=-30$ V.  The ratio $G_\mathrm{exp}/G$ (right axis) as determined using Eq.~(\ref{equ.1}) is shown by open symbols.  Further applying an optimized HV offset $\delta V_{A}$ to Eq.~(\ref{equ.1}) yields the dotted lines.\label{gain_hv}}
\end{figure}

Figure \ref{gain_hv} shows the variation of gain with $\Delta V_A$ for two detectors.
At $\Delta V_{A}=-30$ V we find the gain changes of (4.0~$\pm$~0.7)~\%/V,
where the uncertainty represents one standard deviation.  We found that the observed gain 
can be approximately predicted in terms of $\Delta V_{A}$:

\begin{equation}
 G \propto e^{-|\Delta V_{A}|^{0.361}}.
\label{equ.1}
\end{equation}

\noindent The result of dividing the experimental gain $G_\mathrm{exp}$ by $G$ is shown in Fig. \ref{gain_hv}
by the open symbols. While this empirical parameterization is not perfect for all detectors,
it allows one to predict the gain by better than a factor of two over the range where
the gains change by about a factor of 30.

If one allows for an offset $\delta V_{A}$ in $\Delta V_{A}$, 
$\Delta V_{A}^\prime = \Delta V_{A} + \delta V_{A}$,
then one can obtain a nearly perfect gain prediction, as can be seen by observing
the dotted lines in Fig.~\ref{gain_hv}, which show the experimental gain divided
by $G(\Delta V_{A}^\prime)$. 

Based on slow warm-up tests of the BGO/APD detectors between 90~K and 110 K, we found
that the signals decreased by between 5~\%/K and 8 \%/K
and the breakdown voltage  increased by between 1~V/K and 1.5~V/K
for the group of detectors employed.   
The observed gain changes originate mainly from the temperature dependent APD gains,
as the BGO light output decreases $<$~1~\%/K near 100 K~\cite{Gironnet08}.

A variation in gain was observed with magnetic field for the
field perpendicular to the plane of the APD.  For the BGO/APD detectors at 80~K
the gain for tests with the 662 keV line from the $^{137}$Cs source
varied by $-$5~\% to 30~\%, depending on detector, when the magnetic field was
increased from zero to 4.6~T.  The gain shift was roughly proportional
to the magnetic field. There was no obvious 
correlation between the sign or magnitude of the gain shift  
and the orientation of the detectors around the magnet's axis
(the angle increases in steps of 30 degrees with increasing detector number).

\subsection{APD gain recovery delay after HV cycling}
\label{subsec:gainrecovery}
 
At cryogenic temperatures we found that turning the APD HV off and
immediately back on (\textquotedblleft cycling the HV\textquotedblright) did not restore the gain immediately,
as is the case at room temperature. Indeed, it may take days for the gain
to fully recover even after  shut down for only one second.  

\begin{figure} [p]
\centering
\includegraphics[scale=.50]{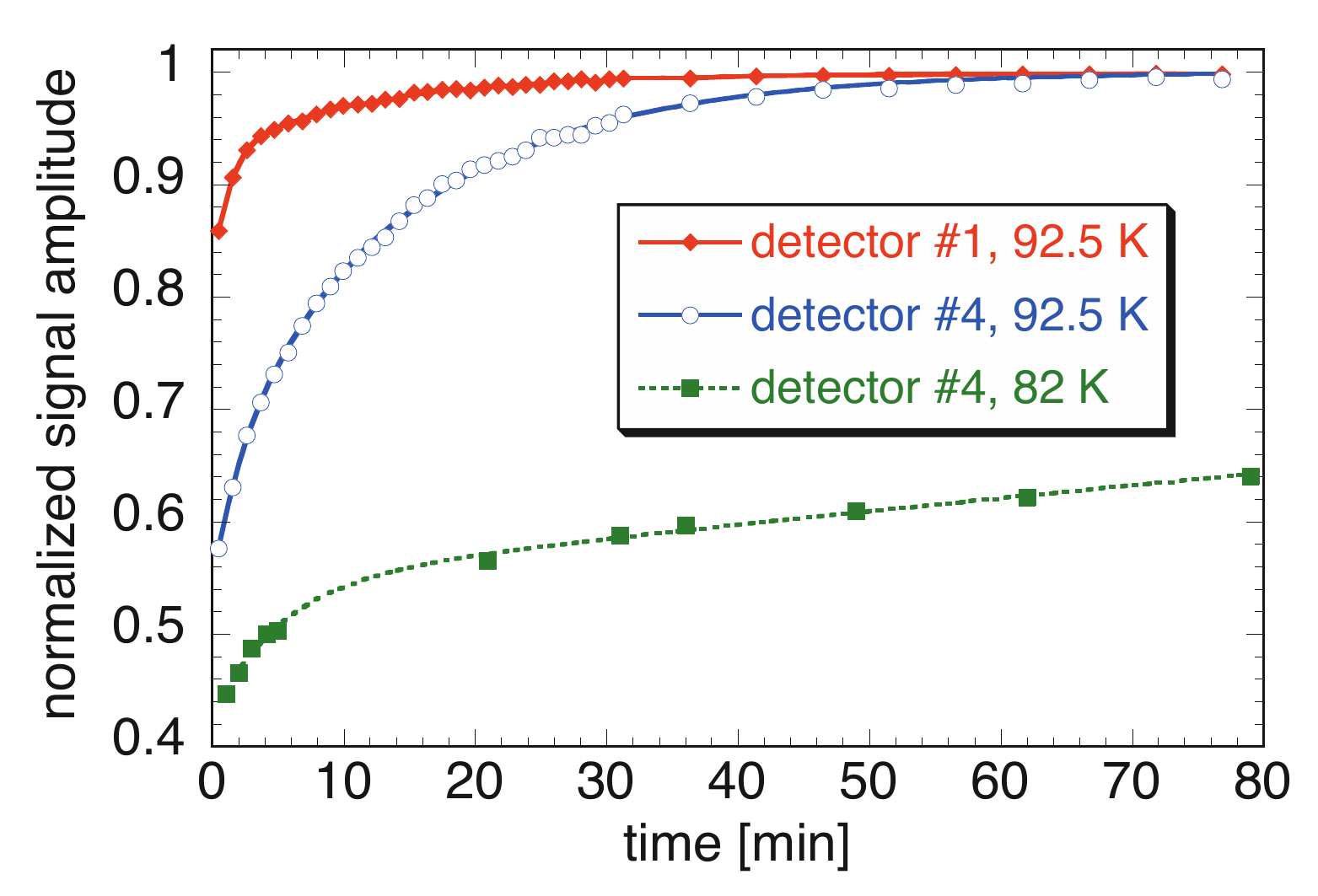}
\caption{The relative APD gain is plotted as a function of time after the 
APD bias has been turned off briefly.  For one APD (labelled detector \#4)
data are shown for 92.5 K (open blue circles) and 82~K (solid green squares).
Four other APDs behaved similarly in tests at 92.5 K, but we also show data for one
detector (labelled detector \#1, red diamonds) for which the effect was less pronounced.
The data are based on fits to the spectra of 662 keV and 511 keV gamma rays.
The solid (92.5 K) and dotted (82 K) lines show fits to a double exponential
parametrization discussed in the text.\label{gain_HVonoff}}
\end{figure}

Fig. \ref{gain_HVonoff} demonstrates this effect for two detectors,
labeled detector \#1 and detector \#4,  at a temperature of 92.5 K.
Four other APDs from the same purchase behaved similarly to detector \#4, but we also show data for 
detector \#1, for which the effect was less pronounced, that was from a later purchase.
The data are normalized to unit signal amplitude at long times after HV cycling.

The data for all APDs tested were well fit by a double exponential form given by
$1- a_S \exp(-t/\tau_S) - a_L \exp(-t/\tau_L)$, where $a_S$, $a_L$, $\tau_S$, and $\tau_L$ are 
adjustable parameters.  Typical values (see Table~\ref{table:gainrecov})
were between 1.6 min and 2.6 min for the short time constant $\tau_S$
and between 14 min and 17 min for the long time constant $\tau_L$.  

\begin{table}[p]
  \begin{tabular} {ccc}
    \hline
    temperature & $\tau_S$ & $\tau_L$ \\ 
    \hline
    82 K & 5 min & 330 min \\
    93 K & (1.6~to~2.6) min  & (14~to~17) min  \\
    115 K & - & $<$ 5 min \\
    \hline
  \end{tabular}
  \caption {Fitted time constants $\tau_S$ and $\tau_L$ for the double exponential form employed to
  parametrize the observed data for recovery of the APD gain after cycling the HV.
  The time scale observed for gain recovery depends strongly on temperature.}
  \label{table:gainrecov}
\end{table}

The pre-power-cycling gain value and several-hour-delayed  \textquotedblleft steady-state\textquotedblright\
gain values differ by nearly two percent. The origin of this
gain change has not been investigated; it may indicate a separate very-long-term 
gain drift or a hysteresis effect from having turned on the HV originally
at a different temperature.

The gain recovery time constant depends strongly on the temperature of
the APDs. This is apparent from the very slowly rising gain recovery curve taken at
82 K, also shown in Fig.~\ref {gain_HVonoff}.  At this temperature we obtained $\tau_S$=5 min
and $\tau_L$=330 min, but we note that the double exponential form is purely phenomenological
and we did not take data for the many hours that would be required to follow the recovery at 82 K.
In contrast, at 115~K (data not shown) $\tau_L$ was measured to be less than 5 min.

The origin of these effects is not understood, but could be due to increased
time to establish the depletion region upon cycling the HV.
As expected, the gain recovery effect was also observed in the larger APDs for direct detection
of 5.9 keV X-rays from the $^{55}$Fe source.
Hence the effect is apparent for both optical and direct soft X-ray detection.
During the radiative decay experiment the bias for the APDs
was turned on before cooling the APDs and left on for the entire time
the apparatus was at cryogenic temperatures.
It is possible that not far below 80~K the APDs become unusable after HV cycling
due to months-long recovery times.  However, applying HV first and then cooling the
detectors may be feasible. It is noteworthy that, at the temperatures studied, reducing the HV by
only a few hundred volts did not show the delay in gain recovery. This could imply that
after cooling to well below 82~K with the HV already applied at higher
temperatures it may still be possible to vary the APD gain by changing the HV.

\section{Summary}
\label{sec:summary}

For an experiment to detect the photons that accompany neutron beta decay,
a compact photon detector  was constructed that could operate in
high magnetic field and near liquid nitrogen temperature.
The scintillator/APD detector operates from a low energy 
threshold of $\approx$~5 keV to $\approx$~1000 keV. 
A large solid angle for detection was obtained by using long BGO crystals
assembled in a tight cylindrical configuration.    The scintillation light was read out
by APDs.  The low energy detection threshold was achieved by operating the
BGO crystals and the APDs near liquid nitrogen temperatures, by minimizing the amount of detector-covering materials,
and by minimizing noise pickup.
BGO light extraction studies found improved performance by selectively applying diffuse reflective paint.
Large sensitivity gains may be possible by better optical coupling of APDs to BGOs,
but no suitable cryogenic coupling agent was found.
The mechanical design optimizes the solid angle coverage while fitting into the
limited space (117 mm diameter) of the bore of a superconducting solenoidal magnet without compromising 
the capability of precise detector alignment around an axial neutron beam.
The detection range was increased to include 0.2 keV to 20 keV by directly
detecting soft X-rays with APDs. 

During testing and operation of the detector system several noteworthy properties of the APD
detectors, relevant at cryogenic temperatures, became apparent which may be important to
consider when planning cryogenic APD operations.  Besides the known properties
of strong sensitivity to temperature and bias voltage, we observed pronounced delays
in achieving steady-state gains when cycling the bias; this delay rapidly increases with
falling temperature and may approach several months below about 70~K.

\vspace{0.3 in}

{\bf ACKNOWLEDGEMENTS} \\

We thank R. Farrell for numerous discussions about APD operation and properties
and L.T. Hudson and R.M. Lindstrom for critical reading of the manuscript.
This work was supported in part by the Department of Energy and the National Science Foundation.

\pagebreak

\bibliographystyle{elsarticle-num}
\bibliography{RDKRef}

\end{document}